\documentclass[twocolumn,showpacs,superscriptaddress,showkeys,reprint,amsmath,amssymb,aps,prd]{revtex4-1}
\usepackage{graphicx}
\usepackage{dcolumn}
\usepackage{bm}
\usepackage{tabularx}
\usepackage{amsmath}
\usepackage{enumitem}
\usepackage{xspace}
\usepackage[bookmarks,bookmarksnumbered]{hyperref}
\hypersetup{colorlinks = true,
            linkcolor = blue,
            anchorcolor =red,
            citecolor = blue,
            pdfauthor=author}
\usepackage[capitalise]{cleveref}

\begin{document}

\newcommand{\UCAS}{School of Physical Sciences, University of Chinese Academy of Sciences (UCAS), Beijing 100049, China\xspace}
\newcommand{\IHEP}{Institute of High Energy Physics, Chinese Academy of Sciences, Beijing 100049, China\xspace}
\newcommand{\ls}{L{\"u}scher\xspace}
\newcommand{\LF}{L{\"u}scher's formula\xspace}
\newcommand{\NN}{neural network\xspace}

\title{Rediscovery of Numerical L{\"u}scher's Formula from the Neural Network}

\author{Yu Lu}
\email{ylu@ucas.ac.cn}
\affiliation{\UCAS}

\author{Yi-Jia Wang}
\email{wangyijia18@mails.ucas.ac.cn}
\affiliation{\UCAS}

\author{Ying Chen}
\email{cheny@ihep.ac.cn, corresponding author}
\affiliation{\UCAS}\affiliation{\IHEP}

\author{Jia-Jun Wu}
\email{wujiajun@ucas.ac.cn, corresponding author}
\affiliation{\UCAS}

\begin{abstract}

We present that by predicting the spectrum in discrete space from the phase shift in continuous space, the \NN can remarkably reproduce the 
numerical \LF to a high precision.
The model-independent property of the \LF is naturally realized by the generalizability of the \NN.
This exhibits the great potential of the \NN to extract model-independent relation between model-dependent quantities, 
and this data-driven approach could greatly facilitate the discovery of the physical principles underneath the intricate data.

\end{abstract}

\maketitle

%%%%%%%%%%%%%%%%%%%%%%%%%%%%%%%%%%%%%%%%%%%%%%%%%%%%%%%%%%%%%%%%%%%%%%%%%%%%%%%%%%%%

%
\section{Introduction}
Physicists are always going after a concise description of data. 
Generally, this concise description boils down to analytic expressions or conserved quantities, which are usually 
dodging and hiding and cannot be trapped easily. 
Nowadays the rapid progress of machine learning (ML) techniques are helping physicists to meet their goals, as manifested by the applications, such as AI Feynman\cite{udrescu:2020a, udrescu:2020b} 
and AI Poincar\'e \cite{Liu:2020omw, liu:2021}. 
For a review of ML techniques in physics, see \cite{Carleo:2019ptp} and several applications in hadron physics in Refs.~\cite{Shanahan:2018vcv, Chen:2021giw, Sombillo:2021rxv, Liu:2022plj, Zhang:2022uqk, Chen:2022shj, Liu:2022uex} 
and references therein.

In most cases of modern physics, a concise description is generally realized at more abstract levels, 
such as analytic differential or integral equations whose solutions are supposed to explain the data. 
If these equations are explicitly known but cannot be solved easily even through numerical methods, ML may help to work out the solutions through Physics-informed-neural-network (PINN) approach\cite{raissi:2019}.
In a more challenging case that there are conceptually links between physical principles and realistic phenomena but we cannot write down the exact expressions, maybe we can also resort to the data driven ML for uncovering the underneath connections. 

A typical example is the study of the strong interaction in the low energy regime.
It is known that the properties of hadrons are necessarily dictated by quantum chromodynmics (QCD), the fundamental theory of the strong interactions. 
However, due to the unique self-interacting properties of gluons, the strong coupling constant is large at the low energy regime and makes the standard perturbation theory inapplicable. 
Up to now, lattice QCD (LQCD) is the most important {\it ab initio} non-perturbative method for investigating the low energy properties of the strong interactions. 
LQCD is defined on the discretized Euclidean spacetime lattice and adopts the numerical simulation as its major approach. 
The major observables of LQCD are energies and matrix elements of hadron systems. 
However, except for the properties of ground state hadrons without strong decays, it is usually non-trivial for lattice results (on the Euclidean spacetime lattice) to be connected with experimental observables in the continuum Minkowski spacetime. 
For example, most hadrons are resonances observed in the invariant mass spectrum of multi-hadron system in decay or scattering processes, while what the lattice QCD can calculate are the discretized energy levels of related hadron systems on finite lattices.
Therefore, the connection must be established. 

%%%%%%%%%%%%%%%%%%%%%%%%%%%%%%%%%%%%%%%%%%%%%%%%%%%%%%%%%%
\begin{figure}[thbp]
  \centering
  \includegraphics[width=0.45\textwidth]{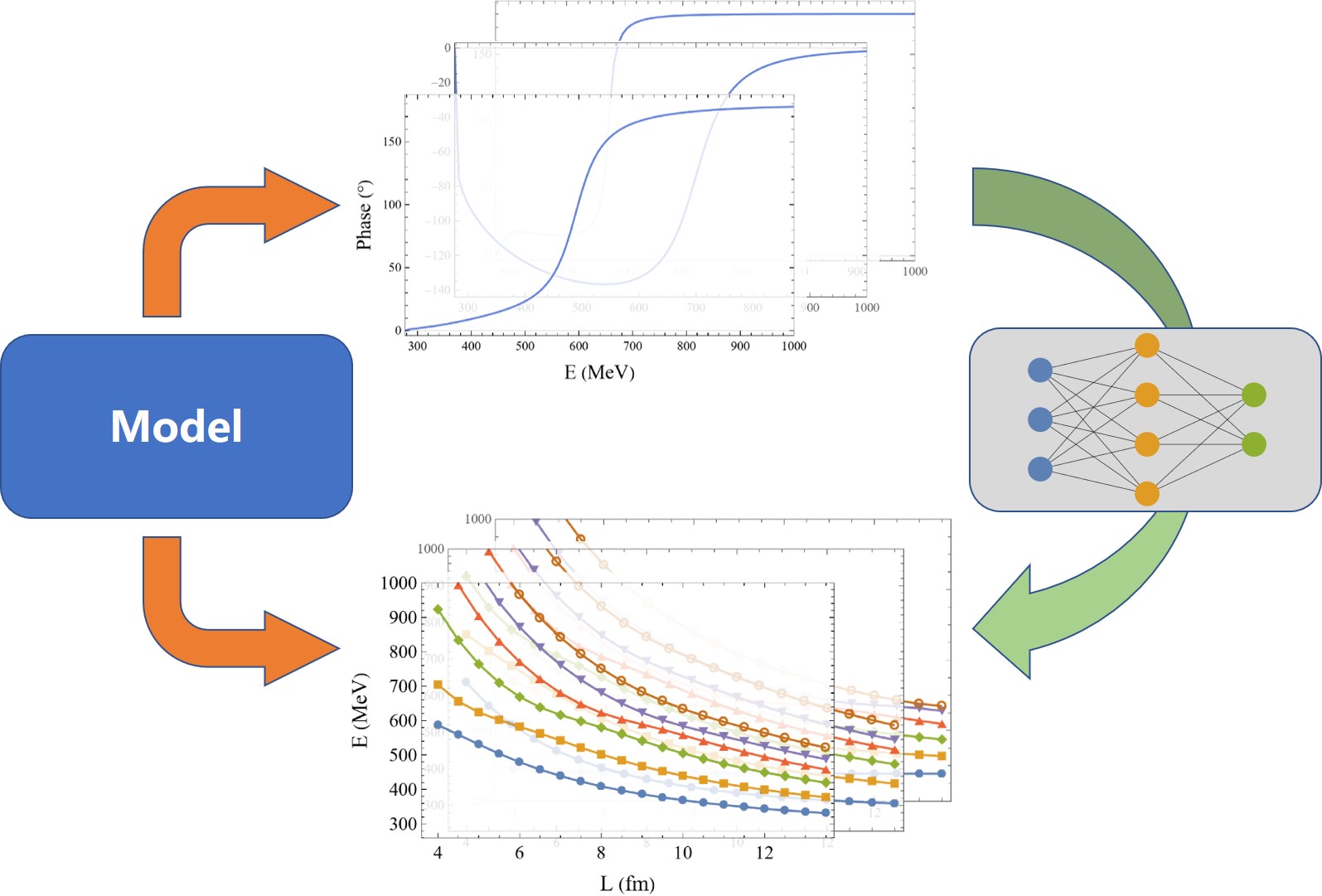} 
  \caption{The workflow of this work.}
  \label{fig:Framework}
\end{figure}
%%%%%%%%%%%%%%%%%%%%%%%%%%%%%%%%%%%%%%%%%%%%%%%%%%%%%%%%%%

One successful approach to address this issue is called \LF \cite{Luscher:1985dn, Luscher:1986pf, Luscher:1990ux}, 
developed by L\"uscher and collaborators more than 30 years ago.
By making use of the finite volume effects, \LF describes relation of the spectrum $E(L)$ of a two-body system on the finite lattice of size $L$ with the scattering phase shift $\delta(E)$ of this system in the continuum Minkowski space. 
The extension of \LF to three-body systems is still undergoing ~\cite{Hansen:2014eka, Jackura:2019bmu, Hansen:2019nir, Blanton:2020gmf, Hammer:2017uqm, Muller:2020wjo, Muller:2021uur,
Mai:2017bge, Mai:2018djl, Brett:2021wyd, Mai:2021lwb}. 
\LF and its extension are not only practically useful, but also is invaluably model-independent on the theoretical side.
Deriving these model-independent theoretical approaches are very challenging and require a lot of wisdom and insight.

For the multi-channel case, more than one free parameters in the scattering amplitude will show up in the infinite volume,
while \LF offers only one constrain to connect the volume size and scattering amplitude at the discrete energy levels.
To extract the information of scattering amplitude in the infinite volume, we will need several different finite volume sizes which share the same energy level.
However, one cannot know a priori which volume size will produce the desired spectra without doing the expensive lattice calculations.
Therefore, one practical way in the multi-channel process is to build a model to relate the scattering amplitude at different energy levels in order to use the \LF to translate the lattice spectrum into phase shifts or other information. 
As a consequence, model dependence will inevitably enter into such calculation.
In contrast, since the \NN is trained via the data-driven way, it is naturally model independent, or at least its model dependence can be safely ignored.
To this end, as a firs step, we should answer whether the \NN can rediscover the numerical \LF in single channel case.

Another challenge comes from the \LF itself.
Unlike extracting the analytic expression of the conserved quantities from the trajectory by ML approach~\cite{liu:2021}, 
the \LF is beyond the elementary function, therefore it can only be evaluated numerically.
This property also leaves a great challenge for \NN to discovery it.

%%%%%%%%%%%%%%%%%%%%%%%%%%%%%%%%%%%%%%%%%%%%%%%%%%%%%%%%%%
\begin{figure}[thbp]
  \centering
  \includegraphics[width=0.45\textwidth]{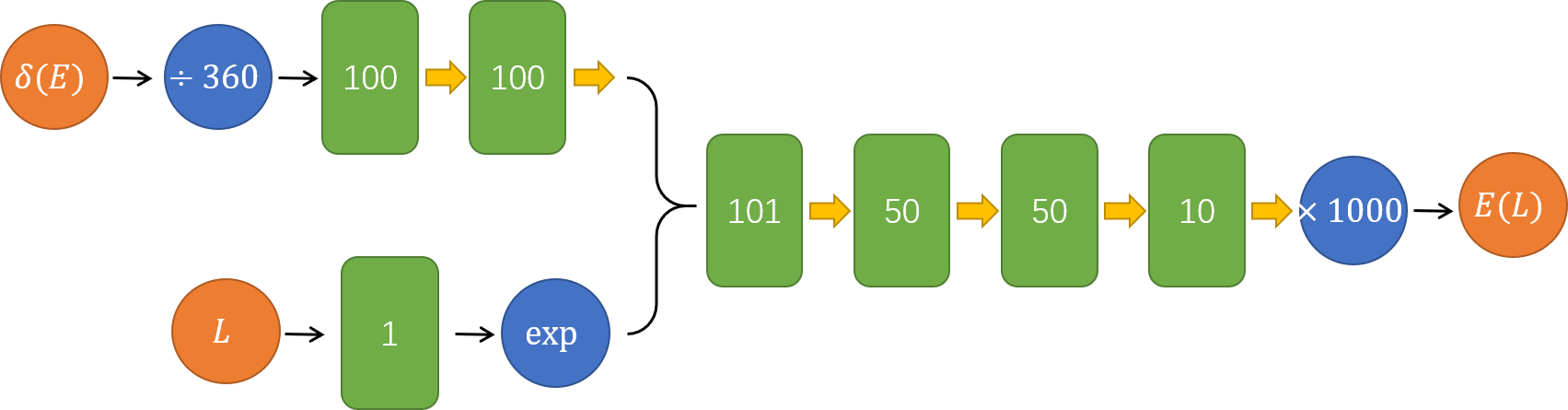}
  \caption{The structure of our neural network.
  Green round rectangles with integer $n$ represent the linear layer with size $n$, which consists of all the learning parameters.
  Orange circles denote the input and output nodes and blue circles are layers with operations marked in the middle.
  The yellow thick arrow marks the ``SoftPlus" activation function and the right brace is a conjunction of the corresponding layers.}
  \label{fig:Framework-NN}
\end{figure}
%%%%%%%%%%%%%%%%%%%%%%%%%%%%%%%%%%%%%%%%%%%%%%%%%%%%%%%%%%

Encouraged by the achievement of machine learning in various areas, 
it is intriguing to ask if the \NN is able to discover the \LF and its variance after fed by plenty data of spectra on lattice and the corresponding phase shifts.
If a model-independence link does exist, in principle, a highly generalizable \NN will be a decent approximation of this link, because of the universal approximation theorem~\cite{hornik:1989, hornik:1991, leshno1993}.
In this paper, we will show that \NN is able to rediscover the numerical \LF to a high precision.

This paper is organized as follows.
Sec.~\ref{sec:formalism} is devoted to build the theoretical formalism to generate the data of energy levels in the finite volume and phase shifts in the infinite space.
Then we will elaborate on the construction of \NN and its training setup in Sec.~\ref{sec:nn}.
In Sec.~\ref{sec:anlay}, we will analyse the result in detail, and provide the evidence to show the numerical form of \LF is generated from \NN.
Finally, we give a brief conclusion in Sec.~\ref{sec:conclusion}.
%

%%%%%%%%%%%%%%%%%%%%%%%%%%%%%%%%%%%%%%%%%%%%%%%%%%%%%%%%%%%%%%%%%%%%%%%%%%%%%%%%%%%%
%%%%%%%%%%%%%%%%%%%%%%%%%%%%%%%%%%%%%%%%%%%%%%%%%%%%%%%%%%%%%%%%%%%%%%%%%%%%%%%%%%%%
%%%%%%%%%%%%%%%%%%%%%%%%%%%%%%%%%%%%%%%%%%%%%%%%%%%%%%%%%%%%%%%%%%%%%%%%%%%%%%%%%%%%

\section{Theoretical Formalism}\label{sec:formalism}
It is known that \LF connects the finite volume energy level $E$ and the $S$-wave phase shift $\delta(E)$  as~\cite{Luscher:1990ux}
\begin{align}
  \delta(E) &= \arctan\left(\frac{q\pi^{3/2}}{\mathcal{Z}_{00}(1;q^2)}\right)+n\pi, \label{eq:luescher}
\end{align}
where $q=\frac{k_{0} L}{2 \pi }$ is defined with $k_0$ being the on-shell momentum of energy $E$, 
and the generalized zeta function $\mathcal{Z}_{00}(1;q^2)$ is defined as
\begin{align}
  \mathcal{Z}_{00} \left(1;q^2\right) &:=\frac{1}{\sqrt{4\pi}}\sum_{\vec{n}\in \mathbb{Z}^3} (\vec{n}^2-q^2)^{-1}.
\end{align}

The system which we check against \LF is the elastic $\pi\pi$ $S$-wave scattering process.
In order to generate the training and test set, 
which consists of the phase shift $\delta(E)$ and the finite volume spectrum $E(L)$ for a given lattice size $L$,
we model this scattering process by Hamiltonian Effective Field Theory (HEFT) ~\cite{Wu:2014vma}.

Following Refs.~\cite{Matsuyama:2006rp,Kamano:2011ih}, 
we assume that $\pi\pi$ scattering can be described by vertex interactions and two-body potentials. 
In the rest frame, the Hamiltonian of a meson-meson system takes the \textit{energy-independent} form as follows,
\begin{eqnarray}
H = H_0 + H_I.
\label{eq:h}
\end{eqnarray}
The non-interacting part is
\begin{eqnarray}
H_0 =|\sigma\rangle m_{\sigma} \langle\sigma|
+ 2\int d\vec{k} |\vec{k}\rangle\omega(|\vec{k}|)\langle\vec{k}|,
\label{eq:h0}
\end{eqnarray}
where $|\sigma\rangle$ is the bare state with mass $ m_{\sigma}$, 
and $|\vec{k}\rangle$ is for the $\pi\pi$ channel state with relative momentum $2\vec{k}$ in the rest frame of $\sigma$, 
and $\omega(k)=\sqrt{m_\pi^2+k^2}$.

The interaction Hamiltonian is
\begin{eqnarray}
H_I = \tilde{g} + \tilde{v},\label{eq:hi}
\end{eqnarray}
where $\tilde{g}$ is a vertex interaction describing the decays of the bare state into two-pion channel,
\begin{eqnarray}
\tilde{g} = \int d\vec{k} \{  |\vec{k}\rangle g^*(k) \langle \sigma| + h.c.\},
\label{eq:int-g}
\end{eqnarray}
and the direct $\pi\pi \to \pi\pi$ interaction (only $S$-wave) is defined by
\begin{eqnarray}
\tilde{v} = \int d\vec{k} d\vec{k}'\, |\vec{k}\rangle  v(k,k') \langle \vec{k}'|.
\label{eq:int-v}
\end{eqnarray}

For the $S$-wave, the $\pi\pi$ scattering amplitude is then defined by the following coupled-channel equation,
\begin{eqnarray}
t(\,k,k'; E)= V(\,k,k')
+\int _0^{\infty} \tilde{k}^{2}d\tilde{k}
\frac{V(k,\tilde{k})t(\tilde{k},k';E)}{E-2\omega(\tilde{k})+i\epsilon}, 
\label{eq:lseq-1}
\end{eqnarray}
where the coupled-channel potential is
\begin{eqnarray}
V(k,k') = \frac{g^*(k)g(k')}{E-m_\sigma} +v(k,k').
\label{eq:lseq-2}
\end{eqnarray}
We choose the normalization
$\langle \vec{k}|\vec{k}^{\,\,'}\rangle  = \delta (\vec{k}-\vec{k}^{\,\,'})$, such that the S-matrix (and thereby the phase shift $\delta(E)$) in each partial-wave is related to the T-matrix by
\begin{eqnarray}
 S(E) \equiv e^{i2\delta(E)} = 1 +2 i T(k_{on},k_{on};E) \label{eq:ST1}
\end{eqnarray}
with
\begin{eqnarray}
T(k_{on},k_{on};E)
=-\pi\frac{k_{on}E}{4}t(k_{on},k_{on};E),\label{eq:ST2}
\end{eqnarray}
and $2\omega(k_{on})=E$. 

On the other hand, the HEFT provides direct access to the multi-particle energy eigenstates in a periodic volume characterized by the size length $L$.
The quantized three momenta of the $\pi$ meson is $k_n = \sqrt{n}\frac{2\pi}{L}$ for
$n = n_x^2+n_y^2+n_z^2$ where $n_x, n_y, n_z=0,\pm1,\pm2, \ldots$.
Then the Hamiltonian matrices with discrete momenta are,
\begin{align}
[H_0]&=\left( \begin{array}{cccccc}
m_\sigma                & 0              & 0                      & \cdots \\
0                       & 2\omega(k_0)   & 0                      & \cdots \\
0                       & 0              & 2\omega(k_1)           & \cdots \\
\vdots                  & \vdots         & \vdots                 & \ddots
\end{array}\right), \\
[H_I]&=\left( \begin{array}{cccccccc}
0                 & \bar{g}(k_0)        & \bar{g}(k_1)       & \cdots \\
\bar{g}(k_0)      & \bar{v}(k_0, k_0)   & \bar{v}(k_0, k_1)  & \cdots \\
\bar{g}(k_1)      & \bar{v}(k_1, k_0)   & \bar{v}(k_1, k_1)  & \cdots \\
\vdots            & \vdots              & \vdots             & \ddots
\end{array} \right).
\label{eq:hi1}
\end{align}
The corresponding finite-volume matrix elements are given by
\begin{align}
\bar{g}(k_n)&=&\sqrt{\frac{C_3(n)}{4\pi}}\left(\frac{2\pi}{L}\right)^{3/2}
g(k_n),\label{eq:gfin}\\
\bar{v}(k_{i},k_{j})&=&\frac{\sqrt{C_3(i)C_3(j)}}{4\pi}\left(\frac{2\pi}{L}\right)^3 v(k_{i},k_{j}),
\label{eq:vfin}
\end{align}
where the factor $C_3(n)$ is the degeneracy of $(n_x, n_y, n_z)$ that gives the same $n$.
The factor $\sqrt{\frac{C_3(n)}{4\pi}}\left(\frac{2\pi}{L}\right)^{3/2}$ follows from the quantization conditions in a finite box of a size $L$, where only $S$-wave contribution is included.
With this Hamiltonian matrix, the spectra in the finite volume are the eigenvalues $E(L)$ of $H$ satisfying $H|\Psi_E\rangle = E(L)|\Psi_E\rangle $.

For $g(k)$ and $v(k,k')$, we let them to be
\begin{eqnarray}
g(k)&=&\frac{g_{\sigma}}{\sqrt{m_\pi}}f(c;k),\label{eq:gepp1}\\
v(k, k')&=&\frac{g_{\pi\pi}}{m^2_\pi}u(d;k)u(d;k')\label{eq:vpp1}
\end{eqnarray}
and in order to explore different types of data, 
three different forms of the $f(a;k)$ and $u(a;k)$ is assumed,
\begin{eqnarray}
f_A(a;k)&=&\sqrt{u_A(a;k)}=\frac{1}{(1+(a k)^2)},\label{eq:formA}\\
f_B(a;k)&=&\sqrt{u_B(a;k)}=\frac{1}{(1+(a k)^2)^2},\label{eq:formB}\\
f_C(a;k)&=&u_C(a;k)=e^{-(ak)^2},\label{eq:formC}
\end{eqnarray}
which are model A, B and C, respectively.

Note that the shapes of the potentials in the momentum space become sharper and sharper from model A to model C.
Since a sharper potential in momentum space has a larger effective range in coordinate space, and therefore has a more prominent finite volume effect, which is an artifact of a finite lattice.
This artifact can be attributed to the deviation of the discrete momentum summation from the continuous momentum integration of the kernel function of the model.
It is proved that the finite volume correction to \LF behaves as $e^{-m L}$ with $m$ being the typical energy scale of the model and a sharper potential will suffer larger corrections in general.

\section{Neural Network And Training Setup}
\label{sec:nn}
To ensure the diversity of the data, the training set should cover both the broadest and the sharpest potentials.
In practice, the data from model A and C are used as the training set, while the data from model B serve as the test set.
Each data set includes 100 evenly-sampled phase shift $\delta(E)$ values from $2m_\pi \approx 277\mathrm{MeV}$ to $1\mathrm{GeV}$,
and the lowest 10 energy levels of $E(L)$ for different lattice sizes $L\in [10,13]$ fm with a step size $0.5$~fm.
The parameter space is spanned by $m_\sigma(\mathrm{MeV}),g_{\sigma},c(\mathrm{fm}),g_{\pi \pi}$ and $d(\mathrm{fm})$, 
ranging from $[350,700]$, $[0.5,5]$, $[0.5,2]$, $[0.1,1]$, $[0.5,2]$, respectively.
The space is randomly sampled by 2500 points for each model A, B and C.
Once the parameters are fixed, 
we calculate the phase shift $\delta(E)$ in continuum space and $E(L)$ for lattice size $L$ in $[10, 13]$~fm with step size $0.5$~fm.
The phase shift $\delta(E)$ is evenly sampled by $100$ points from $2m_\pi \approx 277$~MeV to $1$~GeV, 
and the lowest 10 energy levels of $E(L)$ were kept for training and testing.

We summarize the workflow and the structure of the \NN in Fig.~\ref{fig:Framework-NN}.
Since the phase shift $\delta(E)$ contains the full information of a scattering process, it is natural to expect that the finite volume energy $E(L)$ can be predicted from $\delta(E)$. 
This is treated as a feature extraction task.
To be precise, for a given phase shift $\delta(E)$ and a lattice size $L$, the \NN is designed to predict the \textit{lowest} 10 energy levels $E(L)$ above the $\pi\pi$ threshold.
With some trial-and-error, we construct a small feed-forward fully-connected neural network with ``SoftPlus" activation function.
The \NN is trained by ADAM method, with learning rate $10^{-3}$ and batch size $10^4$.
$10\%$ of the training set is kept for validation and training process finishes after $4\times 10^4$ epoch.

%%%%%%%%%%%%%%%%%%%%%%%%%%%%%%%%%%%%%%%%
\begin{figure}[htbp]
  \centering
      \includegraphics[width=0.45\textwidth]{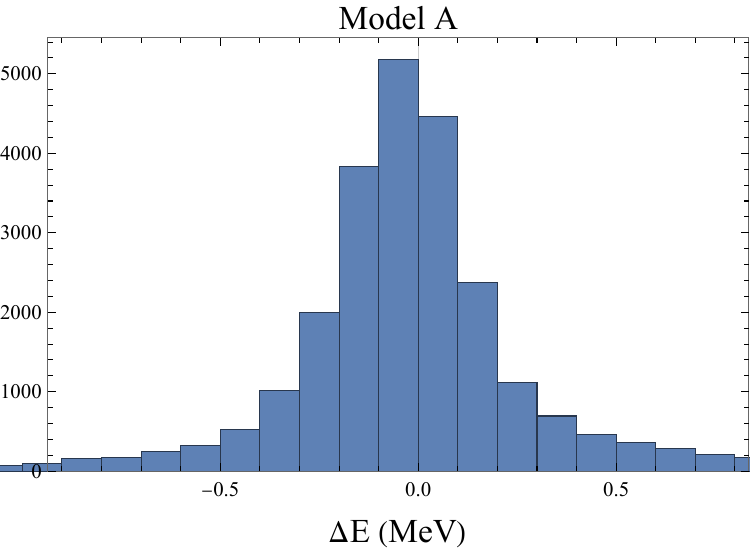}
      \includegraphics[width=0.45\textwidth]{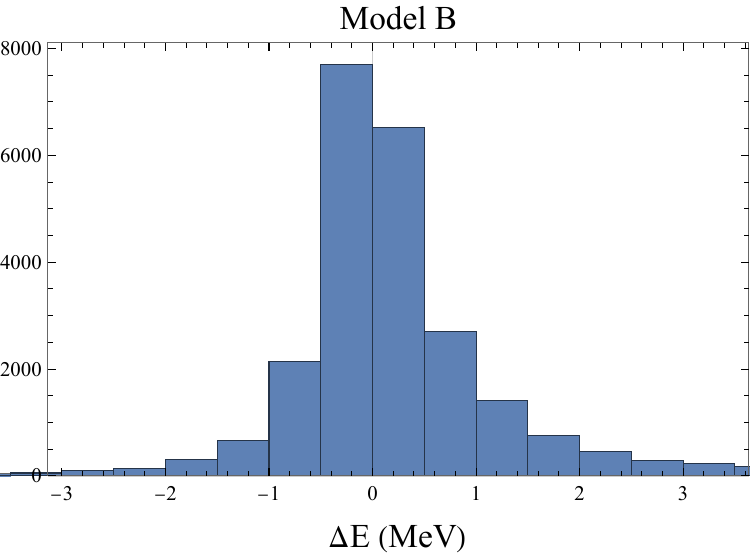}
      \includegraphics[width=0.45\textwidth]{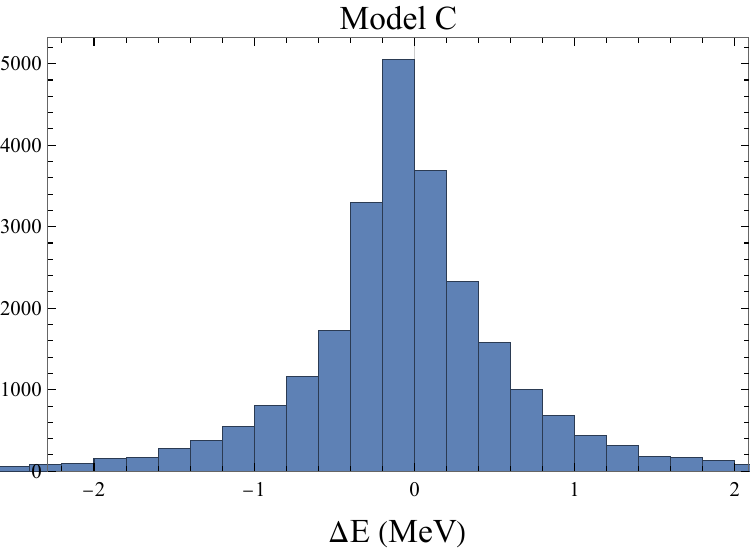}
  \caption{
    The histogram of $\Delta(E)\equiv E_{\mathrm{model}} - E_{\mathrm{NN}}$ at $L=10$~fm, where $E_{\mathrm{model}}, E_{\mathrm{NN}}$ represent the predictions from the \NN and the model, respectively. 
    The \NN is trained on the data from model A and C, and the data from model B serves as test set.
  }
  \label{fig:Histogram}
\end{figure}
%%%%%%%%%%%%%%%%%%%%%%%%%%%%%%%%%%%%%%%%

In Fig.~\ref{fig:Framework-NN}, $L$ and $\delta(E)$ are fed separately into different ports, since they are different physical quantities ($\delta(E)$ has nothing to do with $L$).
In order to speed up the training, we also normalize the input $\delta(E)$ by dividing $360$ and output $E(L)$ by multiplying $1000$.
Instead of the widely used rectified linear unit, we also find our task prefers more smooth activation functions, such as ``SoftPlus".
Although a strict proof is still missing, we speculate this preference of smooth activation functions originates from the regularity or even analyticity of the formulas in physics.
Compared with the tiny \NN such as LeNet-5 in computer vision\cite{LeNet}, our network is even smaller.
However, it turns out that such a simple network is already adequate to make notable predictions.

%%%%%%%%%%%%%%%%%%%%%%%%%%%%%%%%%%%%%%%%%%%%%%%%%%%%%%%%%%%%%%%%%%%%%%%%%%%%%%%%%%%%
%%%%%%%%%%%%%%%%%%%%%%%%%%%%%%%%%%%%%%%%%%%%%%%%%%%%%%%%%%%%%%%%%%%%%%%%%%%%%%%%%%%%
%%%%%%%%%%%%%%%%%%%%%%%%%%%%%%%%%%%%%%%%%%%%%%%%%%%%%%%%%%%%%%%%%%%%%%%%%%%%%%%%%%%%

\section{Results Analysis}
\label{sec:anlay}
For a regression task, one natural test is to calculate the deviation $\Delta(E)=E_{\mathrm{model}}-E_{\mathrm{NN}}$ of the \NN prediction $E_{\mathrm{NN}}$ from the ground truth values $E_{\mathrm{model}}$ from models.

As shown by the histograms in Fig.~\ref{fig:Histogram},
$\Delta(E)$s of all the three models cluster around zero which ensures the precision of the \NN.
It is also reasonable to see the precision on the test set (model B), is slightly worse than that of the training set (model A and C).

\begin{figure*}[htb]
\centering
    \includegraphics[width=0.3\textwidth]{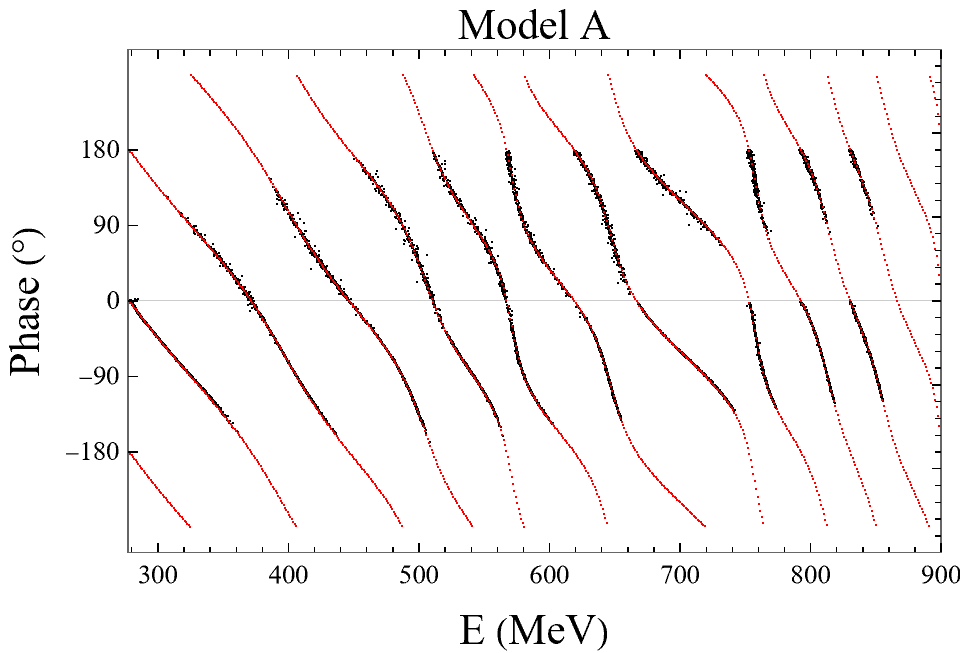}
    \includegraphics[width=0.3\textwidth]{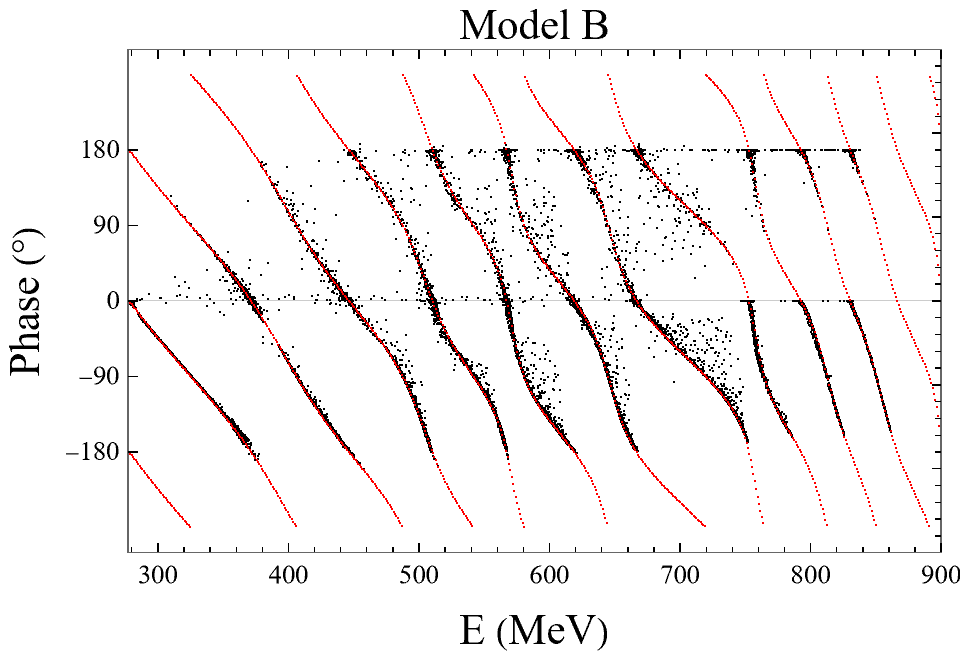}
    \includegraphics[width=0.3\textwidth]{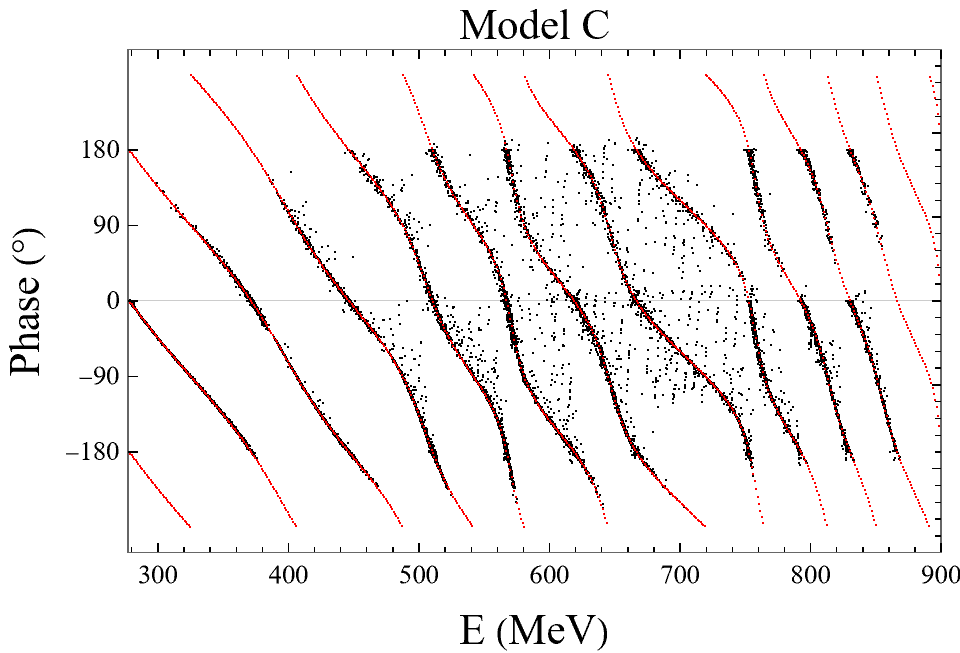}  
\hfill
\centering
    \includegraphics[width=0.3\textwidth]{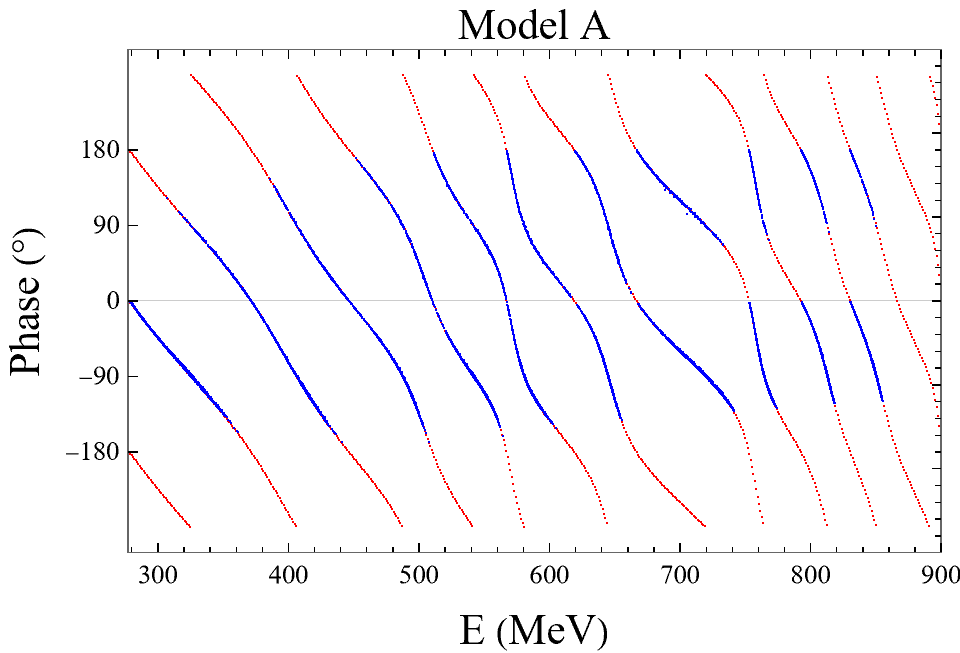}
    \includegraphics[width=0.3\textwidth]{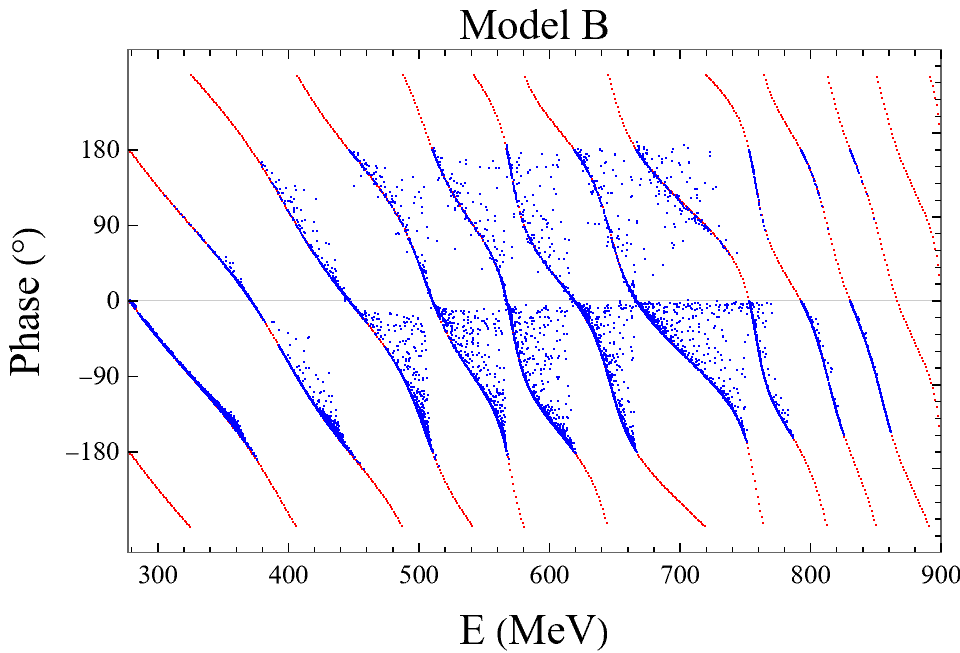}
    \includegraphics[width=0.3\textwidth]{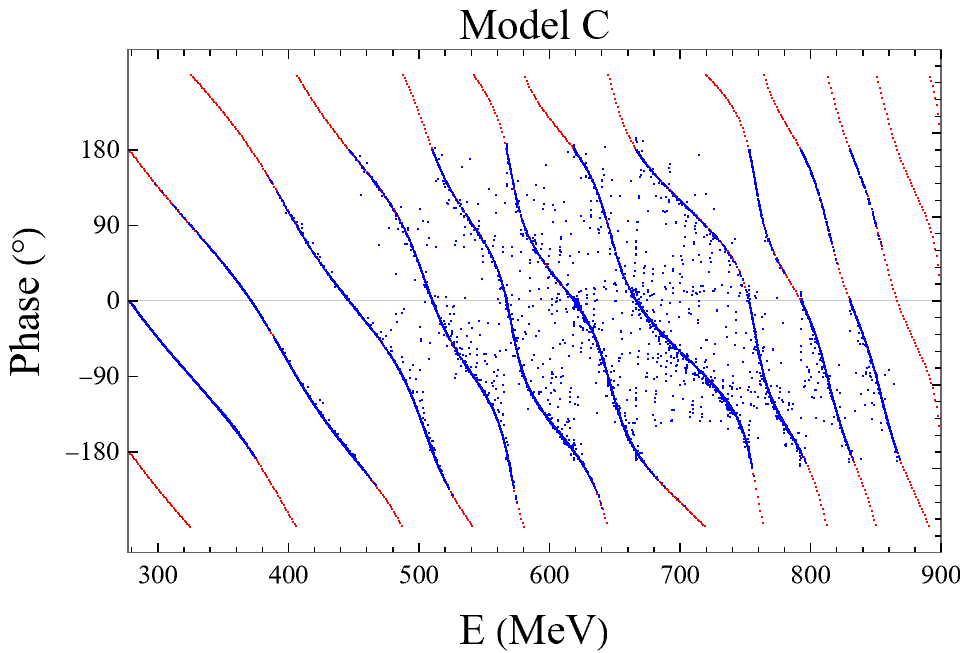} 
\caption{
Comparison of the \LF (red), predictions from the \NN (black) and models (blue), where lattice size is $10$~fm.
}
\label{fig:ScatterTwist}
\end{figure*}

For the test set, there is an additional feature in Fig.~\ref{fig:Histogram}:
The distribution of $\Delta(E)$ has a slightly heavier tail on the right.
This implies that $E_{\mathrm{NN}}$ is generally smaller than $E_{\mathrm{model}}$.
It turns out, this systematic underestimation of the spectrum is not a flaw of \NN, on the contrary, it reveals that \NN is successfully trained as a decent model-independent feature extractor which essentially approximates the \LF.

To see this, we plot the \LF along with the model predictions in Fig.~\ref{fig:ScatterTwist}, where we insert the spectrum back to the phase shift and make a scattering plot of $[E_L, \delta(E_L)]$.

By definition, these points should agree with the \LF up to a \textit{model-dependent} correction term $e^{-mL}$.
Theoretically, it is difficult to foreseen the magnitude or even the sign of this correction term.

\begin{figure}[htb]
\centering
    \includegraphics[width=0.45\textwidth]{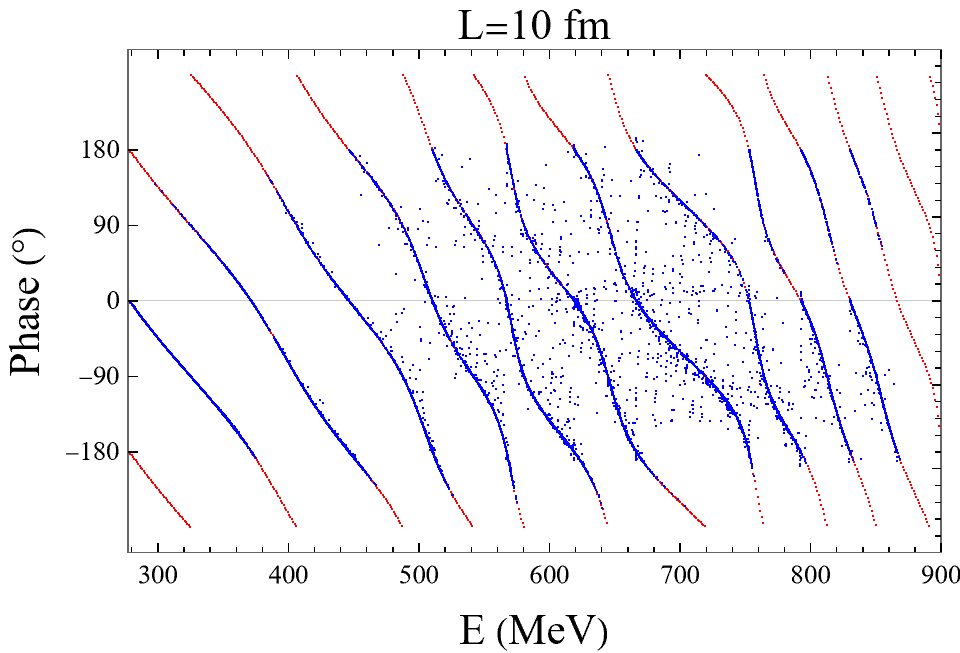}
    \includegraphics[width=0.45\textwidth]{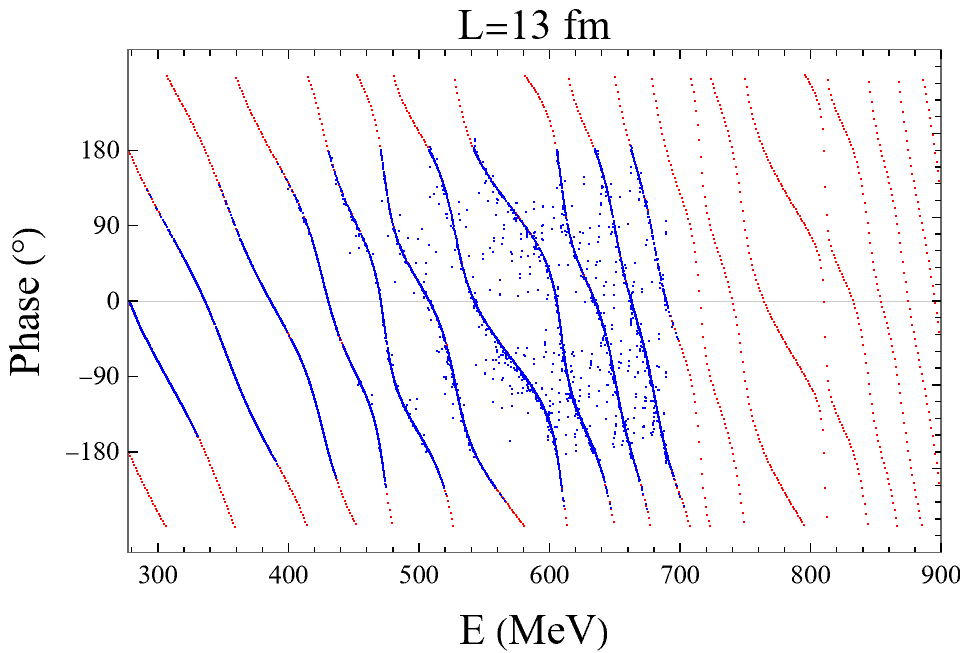}
\caption{
Comparison of model C (blue points) with the \LF in different volume sizes.
}
\label{fig:LFC10-13}
\end{figure}

As shown in Fig.~\ref{fig:ScatterTwist}, data from model A are nearly identical to what \LF predicts. 
This agrees with what we have anticipated, since the potentials in model A are generally narrow in coordinate space, the correction terms are small.
Compared with this ideal-matching case, data from model C becomes much nosier.
However, the mainstream of it still agrees considerably with \LF, 
and the blue points are evenly scattered along two sides of red curves.
It can also be seen in Fig.~\ref{fig:LFC10-13} that, with the increase of the volume size, the model becomes closer to the \LF predictions.

Compared with model A and C, a new feature from the model B is that the spectra from the model are systematically larger than what \LF have predicted. 
Thus, if the \NN learns the \LF well, spectra from model B will be naturally larger than \NN predictions.

The above statement can also be confirmed by comparing the two plots of the second column in Fig.~\ref{fig:ScatterTwist}.
After training, the \NN suppresses the energy levels towards what \LF predicts when it applies on model B, thus leading to a less accurate results and notable non-central distribution in Fig.~\ref{fig:Histogram}.
Thus, the deviation on model B signifies that \NN successfully captures the model-independent ingredients in the process $\delta(E) \to E(L)$ and effectively treats the model-dependent features as noise.

We speculate that this may partially due to the small size of the \NN (28362 parameters V.S. $3.5*10^5$ energy points + $5*10^5$ phase shift points in training set), 
which keeps the \NN from learning or even memorizing the highly model-dependent feature
(see. e.g. Ref.~\cite{ZhangBHRV:16} for the risk when the number of parameters exceeds the number of data points).
Since \LF is the only model-independent approach, this lead to our central conclusion that we get a \NN reprint of the numerical \LF.

% \emph{Further Discussions.}---
To make a stronger evidence that the numerical \LF is learned by \NN, it is necessary to expand the test set and explore the generalizability further, i.e., challenge the \NN by more different types of phase shifts.
This will not only reveal more interesting structure of the \NN, 
but also guide us to spot a subtle deficiency in the above treatment.

One typical pattern of the phase shift $\delta(E)$ in our training and test set is that, with the increase of energy, $\delta(E)$ will departure from zero at $2m_\pi$ threshold, gain a sharp or broad resonance structure in the middle steps and end up to be $0^\circ$ or $\pm 180^\circ$.
Here, we will challenge the \NN by feeding a \textit{constant} phase shift $\delta(E)=\delta_0$, where $\delta_0$ ranges in $[-180^\circ, +180^\circ]$.
Since this constant phase shift is far beyond our training set, it would be impossible to pass the test if the \NN were doing nothing but a trivial memorization.

\begin{figure}[htb]
  \centering
  \includegraphics[width=0.45\textwidth]{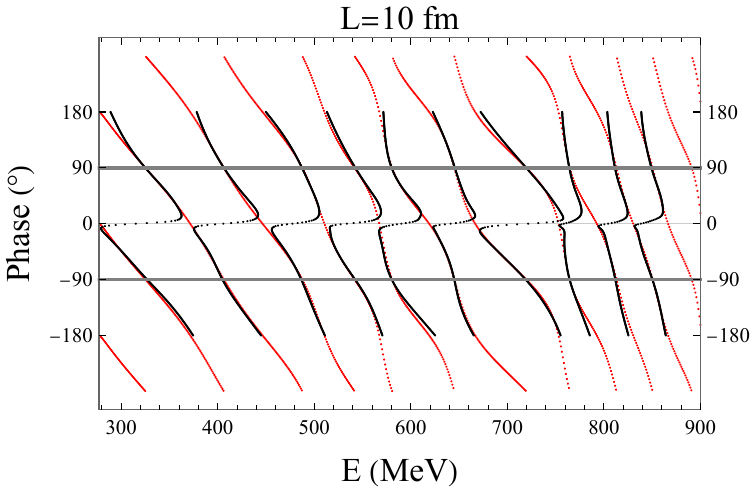}
  \caption{
    Prediction (black dots) from the \NN when phase shift is constant $\delta(E)=\delta_0, \delta_0\in [-180^\circ,180^\circ]$.
    The precise \LF curve is marked as red dots.
    One period boundary $\pm 90^\circ$ is marked by gray horizontal line for comparison.
  }\label{fig:ScanTwist}
\end{figure}

In Fig.~\ref{fig:ScanTwist}, the agreement between \LF and \NN is even more fascinating except an unexpected twist around $\delta=0^\circ$.
To be precise, if we track the lowest level of the spectrum $E_1$,
the \NN concludes from the data that $E_1$ should generally increase with the decrease of $\delta_0$.
However, once $\delta_0$ crosses the zero from above, another lower energy level will emerge.
Thus, as a function of $\delta_0$, $E_1$ is not a continuous function at zero.
This discontinuity is essentially caused by the periodicity of the phase shift: 
$\delta(E)$ and $\delta(E) + n\pi$ corresponds to the same physics.
On the other side, since the \NN is designed to predict the lowest 10 energy levels above the threshold, and the activation functions are continuous in order to do back-propagation in the training process,
the best \NN can achieve is to make a soft transition between the neighbor red curves around $\delta_0=0^\circ$, 
resulting several zigzag tracks in Fig.~\ref{fig:ScanTwist}.
It is also worthy to find that this twist structure does not manifest itself in Fig.~\ref{fig:ScatterTwist},
which makes this constant-phase-shift-test valuable.

\begin{figure}[htb]
  \centering
  \includegraphics[width=0.45\textwidth]{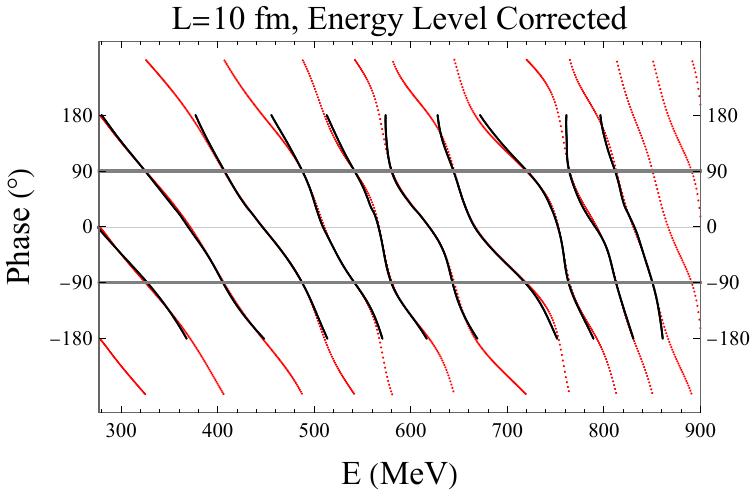}
  \caption{
  Same as Fig.~\ref{fig:ScanTwist}, with energy level issue explicitly addressed.
  }\label{fig:Scan}
\end{figure}

We circumvent this twist issue by the following approach.
The energy level $E$ is marked as $E_1$ only when $\delta(E)$ is negative and $E<2\sqrt{\left(\frac{2\pi}{L}\right)^2 +m_\pi^2}$,
otherwise, the valid energy levels starts from $E_2$.
Noting that this does not request any pre-knowledge of the \LF.
It is essentially a convention that $\delta(E)$ is zero at the following free energies 
\begin{align}
  E_\mathrm{free}:=2\sqrt{\vec{n}^2\left(\frac{2\pi}{L}\right)^2 +m_\pi^2},
  \label{eq:freeE}
\end{align}
where $\vec{n}=(n_x,n_y,n_z), n_{x,y,z}=0,\pm 1,\pm 2,\ldots$.
Retraining the \NN with the above modification results in a superb agreement with the \LF, which is shown in Fig.~\ref{fig:Scan}.
The slightly worse precision around $\pm 180^\circ$ can be improved by either increasing the size of the \NN or we can simply ignore the \NN predictions by constraining it within a period, such as $[-90^\circ,+90^\circ]$ and extrapolate the results to other regions by periodicity.
After addressing this twist issue, we finally strengthen the previous conclusion that the numerical form of \LF is learned by the \NN.

%%%%%%%%%%%%%%%%%%%%%%%%%%%%%%%%%%%%%%%%%%%%%%%%%%%%%%%%%%%%%%%%%%%%%%%%%%%%
%%%%%%%%%%%%%%%%%%%%%%%%%%%%%%%%%%%%%%%%%%%%%%%%%%%%%%%%%%%%%%%%%%%%%%%%%%%%
%%%%%%%%%%%%%%%%%%%%%%%%%%%%%%%%%%%%%%%%%%%%%%%%%%%%%%%%%%%%%%%%%%%%%%%%%%%%

\section{Summary and Outlook}
\label{sec:conclusion}
In this paper, we have shown that the numerical form of \LF can be rediscovered by the \NN when it is trained to predict the spectrum on lattice from the phase shift in continuous space.
From the perspective of pragmatism, the \NN is able to exploit the sophisticated data and extract valuable information.
In a broad sense, our work is an concrete example to demonstrate how to extract model-independent link between model-dependent quantities in a data-driven approach.
Surprised by the capability of the \NN, we believe its potential is still waiting for physicists to explore.

%%%%%%%%%%%%%%%%%%%%%%%%%%%%%%%%%%%%%%%%%%%%%%%%%%%%%%%%%%%%%%%%%%%%%%%%%%%%
%%%%%%%%%%%%%%%%%%%%%%%%%%%%%%%%%%%%%%%%%%%%%%%%%%%%%%%%%%%%%%%%%%%%%%%%%%%%
%%%%%%%%%%%%%%%%%%%%%%%%%%%%%%%%%%%%%%%%%%%%%%%%%%%%%%%%%%%%%%%%%%%%%%%%%%%%

\section{Acknowledgment} 
We are grateful to Yan Li, Qian Wang, Ross D. Young and James M. Zanotti for useful discussions.
This work is partly supported by the National Natural Science Foundation of China (NSFC) under Grants No. 12175239, 12221005, No.11935017, No.12070131001 (CRC 110 by DFG and NNSFC) and by the National Key R$\&$D Program of China under Contract No. 2020YFA0406400. 

%%%%%%%%%%%%%%%%%%%%%%%%%%%%%%%%%%%%%%%%%%%%%%%%%%%%%%%%%%%%%%%%%%%%%%%%%%%%
%%%%%%%%%%%%%%%%%%%%%%%%%%%%%%%%%%%%%%%%%%%%%%%%%%%%%%%%%%%%%%%%%%%%%%%%%%%%
%%%%%%%%%%%%%%%%%%%%%%%%%%%%%%%%%%%%%%%%%%%%%%%%%%%%%%%%%%%%%%%%%%%%%%%%%%%%

\bibliographystyle{elsarticle-num}
\bibliography{NNLuescher}

\end{document}